\documentclass [aps,prl,twocolumn,groupedaddress,showpacs,floatfix]{revtex4}
\usepackage{graphicx}

\begin{document}
\title{Optically tunable surfaces with trapped particles in microcavities}
\author{R.~Sainidou}
\author{F.~J. Garc\'{\i}a de Abajo}
\affiliation{Instituto de \'{O}ptica - CSIC, Serrano 121, 28006
Madrid, Spain}
\date{\today}
\begin{abstract}
We introduce optically tunable surfaces based upon metallic gold
nanoparticles trapped in open, water-filled gold cavities. The
optical properties of the surfaces change dramatically with the
presence and location of the particles inside the cavities. The
precise position of the particles is shown to be controllable
through optical forces exerted by external illumination, thus
leading to all-optical tunability, whereby the optical response of
the surfaces is tuned through externally applied light. We discuss
the performance of the cavity-particle complex in detail and provide
theoretical support for its application as a novel concept of
large-scale optically tunable system.
\end{abstract}

\maketitle

The last decade has witnessed tremendous progress in optical
trapping and manipulation of small particles with a wide range of
applications ranging from physics and engineering to biology and
medicine~\cite{A1997,G03}. One of the most fascinating domains of
those applications is directly related to the possibility of
controlling and tuning the optical response of nanoscale systems,
thus opening the field to bio-sensing and all-optical switching.
Early attempts to bind small particles using light forces
\cite{A1970,AD1975,A1980,AD1987,BFG} led to the development of
optical tweezers \cite{A1970,AD1987}, capable of trapping and
aligning objects ranging from micro-organisms \cite{AD1987} to
metallic nanoparticles \cite{HBH05,PLK06}. Manipulation of
micro-particles using plasmons has been recently demonstrated as
well \cite{RZG07}, whereas fine tuning of nanoparticle positions has
been theoretically proved to be realizable by coupling to plasmonic
nanostructures \cite{NBX97,paper126}.

In this Letter, we investigate the optical tunability performance of
cavity-nanoparticle systems. More precisely, we consider an open
metallic cavity in an otherwise flat gold surface surrounded by
water and containing a gold nanorod in its interior. The
nanoparticle is the key to purposely modify the optical response of
the cavity-particle composite system. More precisely, we demonstrate
through theoretical simulations the possibility of
optically-trapping the particle at positions that depend on the
wavelength of externally incident light, thus producing tunable
changes in the optical response of the system. The driving optical
force can be used to displace the trapped particle at will inside
the cavity, which results in changes of the surface reflectivity.
All calculations presented here, are performed using the
boundary-element method, which we push to full converge within the
scale of the figures~\cite{BEM}.

\begin{figure}[t!]
\centerline{\includegraphics*[width=8.4cm]{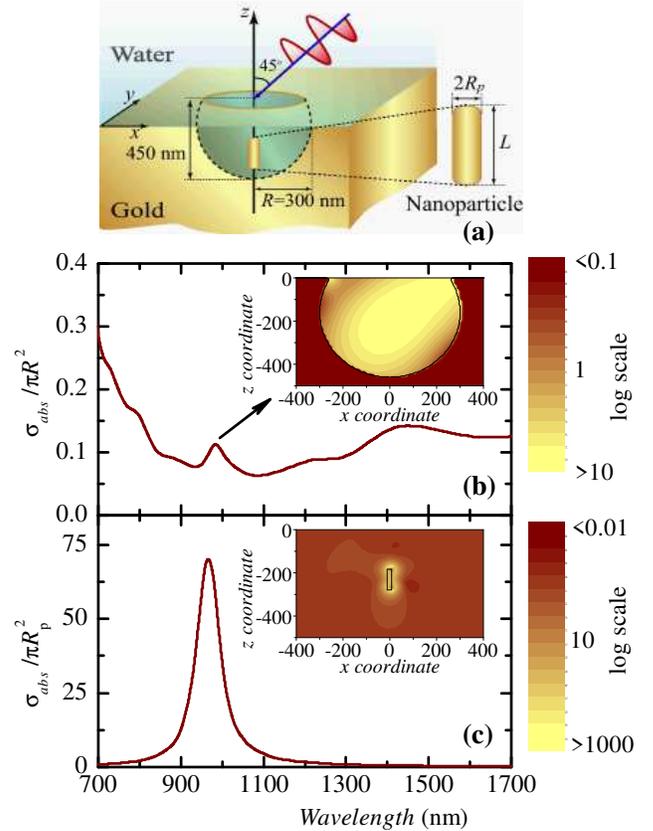}}
\caption{(color online) {\bf (a)} Cavity-particle system under
study, including definitions of the cavity dimensions, and particle
radius $R_p$ and length $L$. The cavity, excavated in an otherwise
infinite planar gold-water interface, is illuminated by
$p$-polarized (TM) light under $45^{\circ}$ incidence with respect
to the planar surface. The gold nanorod is assumed to be placed in
different positions along the cavity axis ($z$ axis). {\bf (b)} and
{\bf (c)} Normalized absorption cross section $\sigma_{\rm abs}$ of
the single cavity and a single nanorod of dimensions
$20~\mathrm{nm}\times100~\mathrm{nm}$, respectively.
Near-field-intensity plots at the resonance wavelengths
$985~\mathrm{nm}$ and $970~\mathrm{nm}$ are shown in the insets.}
\label{fig1}
\end{figure}

The optical properties of spherical nanocavities in planar gold
surfaces are well understood from the experimental and theoretical
points of view \cite{CNB01,paper128}, leading to such phenomena as
omnidirectional total absorption of light \cite{paper150}. We
consider in particular a cavity of radius $R=300~\mathrm{nm}$ and
depth $450~\mathrm{nm}$ immersed in water, as shown in Fig.\
\ref{fig1}(a). The system is illuminated with $p$-polarized light
incident with an angle of $45^{\circ}$ with respect to the flat gold
surface. The cavity exhibits resonances associated to trapped
modes~\cite{paper128}, which appear as sharp peaks in the absorption
spectrum [see Fig.~\ref{fig1}(b)]. In particular, the pronounced
resonance at $985~\mathrm{nm}$ is associated to a near-field
intensity distribution with an important localization at the center
of the cavity [see inset of Fig.~\ref{fig1}(b)], implying the
possibility of strong interaction with the corresponding plasmon
modes of any nanoparticle placed in that region. The nanoparticle
itself will obviously exhibit plasmon modes, which can be tuned
close to the cavity resonance wavelength. Gold nanorods are good
candidates for that purpose, since their longitudinal plasmon modes
can be easily excited~\cite{BH98}. The wavelength position of these
resonance modes depends strongly on the dimensions of the
cylindrical rod, denoted here as $(diameter=2R_p)\times(length=L)$.
An example is given in Fig.~\ref{fig1}(c) for a
$20~\mathrm{nm}\times100~\mathrm{nm}$ gold nanorod. A longitudinal
plasmon is clearly observed at $970~\mathrm{nm}$.

\begin{figure}[t]
\centerline{\includegraphics*[width=8.5cm]{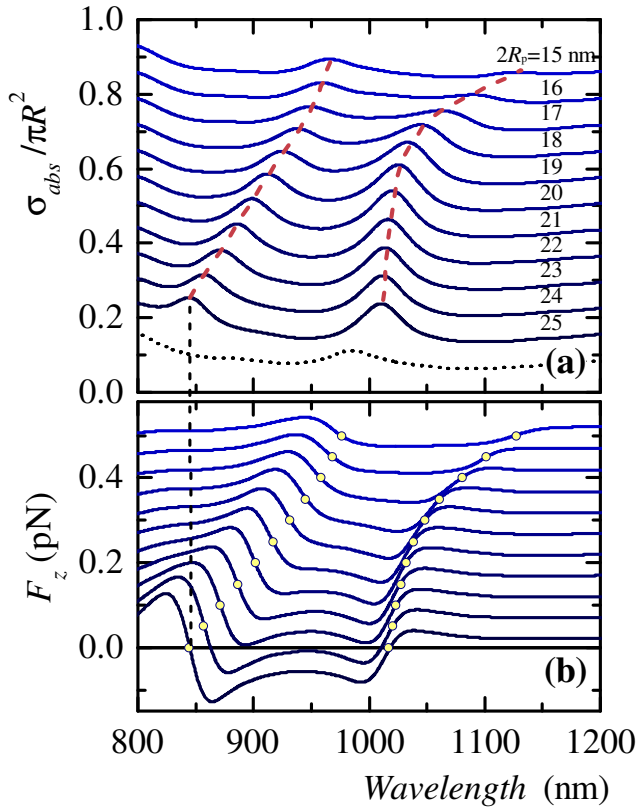}}
\caption{(color online) {\bf (a)} Absorption spectra of the
rod-cavity composite system for different rod diameters $2R_p$ (see
labels) and fixed rod length $L=100~\mathrm{nm}$. The rod is placed
along the $z$ axis with its center $210~\mathrm{nm}$ below the
cavity opening. The spectra are vertically displaced for clarity.
The dashed lines indicate the position of resonance maxima. The
dotted curve shows the cavity without a particle. {\bf (b)} Vertical
force acting on the rod under the same conditions as in (a) for an
external light intensity of $1~\mathrm{mW/\mu m^2}$. The different
curves are displaced for clarity with the open symbols denoting the
zero-force points.} \label{fig2}
\end{figure}

When the particle is placed inside the cavity, the interaction
between their respective plasmon resonances will depend on the
overlap between their associated fields. We will assume for
simplicity that the particle is aligned along the rotational axis of
the cavity (the $z$ axis), and therefore strong overlap will occur
for cavity modes localized at the center of a cavity such as that of
Fig.~\ref{fig1}(b). Let us first assume that the rod is in the
center of the cavity (i.e., $210~\mathrm{nm}$ away from the
opening), close to the maximum of the cavity-mode intensity. The
interaction between cavity and rod gives rise to two distinct peaks
in the absorption spectrum of the composite system that are
displaced with respect to the non-interacting cavity and rod modes.
For instance, we observe modes at wavelengths $910~\mathrm{nm}$ and
$1025~\mathrm{nm}$ for the $20~\mathrm{nm}\times100~\mathrm{nm}$
gold nanorod, far away from the initial resonances of the two
individual constituents of the system, formerly at wavelengths of
$970~\mathrm{nm}$ (rod) and $985~\mathrm{nm}$ (cavity), as shown in
Fig.~\ref{fig1}. The aspect ratio (AR) of the rod plays a crucial
role for the wavelength position of these two resonances, as the
longitudinal mode of the nanorod can be swapped over a wavelength
interval around the cavity mode wavelength for varying AR. This is
illustrated in Fig.~\ref{fig2}(a), where an avoided crossing is
clearly discernible.

\begin{figure}[t]
\centerline{\includegraphics*[width=8.5cm]{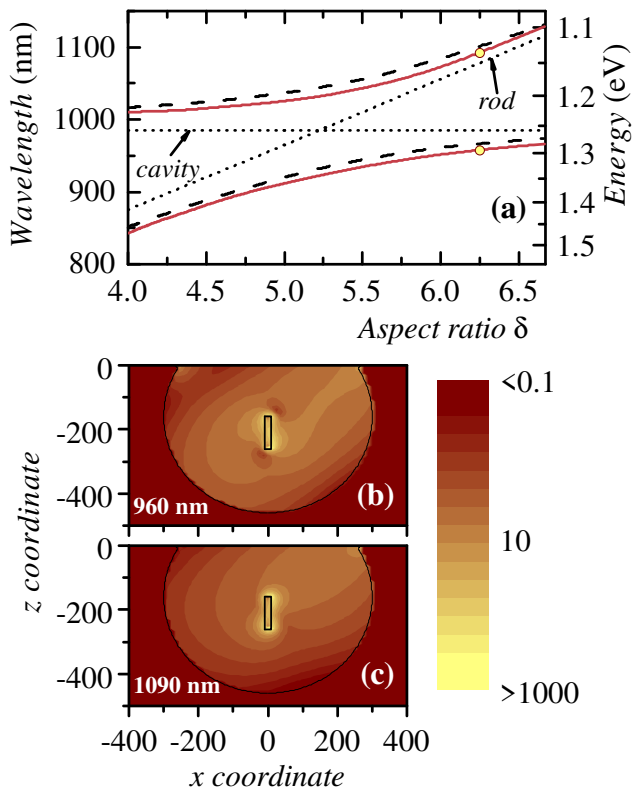}}
\caption{(color online) {\bf (a)} Resonance modes of the rod-cavity
system extracted from Fig.\ \ref{fig2}(a) and represented as a
function of the particle aspect ratio $\delta=L/2R_p$ (solid
curves). Dashed curves: analytical-model approximation [see
Eq.~(\ref{eq:hybr})]. Dotted curves: single-cavity and single-rod
resonance modes. {\bf (b)} and {\bf (c)} Near-field-intensity plots
for the $16~\mathrm{nm}\times100~\mathrm{nm}$ rod [see open circles
in (a)] at the resonance wavelengths of the composite system,
$960~\mathrm{nm}$ and $1090~\mathrm{nm}$, respectively.}
\label{fig3}
\end{figure}

In a quantum-mechanical analogy~\cite{M1966}, we can assume that the
cavity-nanoparticle composite system is governed by an effective
Schr\"odinger equation of the form
$\hat{H}|\Phi\rangle=\omega|\Phi\rangle$, where the Hamiltonian
$\hat{H}=\hat{H}_0+\hat{V}$ is the operator of the
cavity-nanoparticle composite; $\hat{H}_0$ is the operator
describing the individual constituents of the system with
eigenstates $|n_1\rangle$ (cavity mode) and $|n_2\rangle$ (rod mode)
and corresponding eigenvalue frequencies $\omega_1$ and $\omega_2$;
$\hat{V}$ describes the interaction between cavity and particle; and
we can approximate the state of the system as
$|\Phi\rangle=a_1|n_1\rangle+a_2|n_2\rangle$, that is, the
hybridization of cavity and particle modes with superposition
coefficients $a_1$ and $a_2$, respectively. We can safely assume a
linear dependence of the hopping parameter $u=\langle
n_1|\hat{V}|n_2 \rangle=\langle n_2|\hat{V}|n_1\rangle$ on the AR of
the rod (i.e. $u=u_1+u_2\delta$), and find that the two
eigenfrequencies $\omega_\pm$ of the composite system are given by
\begin{equation}
\omega_{\pm}=\frac{\omega_1+\omega_2}{2}\pm
\sqrt{\left(\frac{\omega_1-\omega_2}{2}\right)^2+(u_1+u_2\delta)^2}
\;. \label{eq:hybr}
\end{equation}
The quantities $u_1$ and $u_2$ are to be determined by fitting
Eq.~(\ref{eq:hybr}) to the values $\omega_{\pm}$ found numerically
from the absorption spectra of Fig.~\ref{fig2}(a), which are
represented as solid curves in Fig.~\ref{fig3}(a). We find
$u_1=0.146~\mathrm{eV}$ and $u_2=-0.0144~\mathrm{eV}$, and the
analytical values of Eq.~(\ref{eq:hybr}) for $\omega_{\pm}$ using
these parameters are also shown for comparison [dashed curves in
Fig.~\ref{fig3}(a)].

The hybridization can be also observed in the near-field plots at
the resonance wavelengths of the composite system. An example is
given in Figs.~\ref{fig3}(b) and (c) for the
$16~\mathrm{nm}\times100~\mathrm{nm}$ rod. The near-field plots
clearly show a rod-like and a cavity-like profile at the
corresponding resonance wavelengths of $960~\mathrm{nm}$ and
$1090~\mathrm{nm}$ [cf. insets in Figs.~\ref{fig1}(b) and (c)],
since they are located close to the single-rod ($1080~\mathrm{nm}$)
and single-cavity ($985~\mathrm{nm}$) modes, respectively. We intend
to obtain significant changes in the resonance wavelength of the
composite system triggered by the presence of the nanoparticle, and
thus, we need to work close to the crossing point
($\omega_1=\omega_2$) of the single-rod and single-cavity modes
[i.e., with AR $\delta \simeq 5.25$, according to
Fig.~\ref{fig3}(a)] in order to maximize mode repulsion,
approximately given by $\pm|u_1+u_2 \delta|$. Therefore, we will
consider $20~\mathrm{nm}\times100~\mathrm{nm}$ rods in what follows.

The observed change in the optical response of the cavity due to the
presence of the nanorod in its interior cannot be realizable if we
do not have a tool to keep the nanoparticle trapped inside the
cavity at a designated position. In fact, our system offers such a
possibility as we show next. We calculate the optical force exerted
on the particle for a given external illumination following the
methods described for instance in Ref.~\cite{paper089}. Hereafter,
we will only be concerned with the $z$-component force, $F_z$,
assuming that its $x$-component is canceled out by adding a second
external light beam propagating to the opposite $x$-direction with
respect to the first one (i.e., the two beams are symmetrically
placed with respect to the $yz$-plane). We represent in
Fig.~\ref{fig2}(b) the force acting on the nanorods considered in
Fig.~\ref{fig2}(a) for an external light-intensity of
$1~\mathrm{mW/\mu m^2}$. The magnitude of the force is large enough
to overcome effects due to friction or Brownian motion, which
typically contribute with forces of a few
$\mathrm{fN}$~\cite{forces}. Also notice that the force is
proportional to the light intensity, so that even much weaker
incident illumination will produce sufficient trapping stability,
particularly if one is concerned with the response of a
statistically large number of cavities, as it is the case in
macroscopic surfaces \cite{CNB01}.

\begin{figure}[t]
\centerline{\includegraphics*[width=8.5cm]{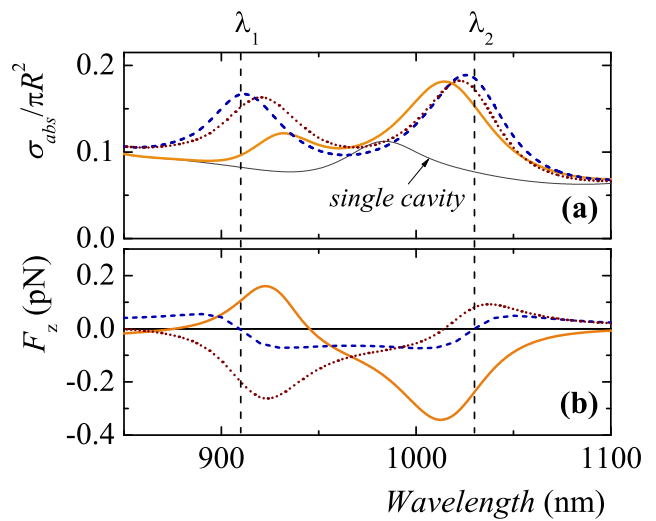}}
\caption{(color online) {\bf (a)} Absorption spectra of the
rod-cavity system for a $20~\mathrm{nm}\times100~\mathrm{nm}$ rod
placed at different depths from the cavity opening:
$-60~\mathrm{nm}$, $-230~\mathrm{nm}$, and $-330~\mathrm{nm}$
(solid, dashed, and dotted curves, respectively). The single cavity
absorption is shown for reference. {\bf (b)} Forces along the $z$
axis acting on the rod for the cases considered in (a). The external
light intensity is $1~\mathrm{mW/\mu m^2}$. The vertical dashed
lines are guides to the eye (see text).} \label{fig4}
\end{figure}

The absorption spectrum of the cavity-rod system is quite sensitive
to changes in the position of the rod along the $z$ axis, as
expected from the degree of overlap between the particle and the
cavity mode intensity [see Fig.~\ref{fig1}(b)]. This is illustrated
in Fig.~\ref{fig4}(a). Our system is thus a good candidate to
exhibit all-optical tunability: the use of optical forces exerted on
the particle through external illumination can displace the nanorod
along the $z$ axis to trigger changes in the absorption/reflection
of the system. We show in Fig.~\ref{fig4}(b) these forces,
calculated for an external light intensity of $1~\mathrm{mW/\mu
m^2}$. Three different rod positions along the $z$ axis are
considered ($60$, $230$ and $330~\mathrm{nm}$ from the cavity
aperture). For the first resonance mode located at lower
wavelengths, the force is directed downwards when the particle lies
at the deepest position (dotted curves), and it changes to upwards
at less deep positions (dashed and solid curves). Considering for
example an operating wavelength $\lambda_1=910~\mathrm{nm}$, the
optical force can remove the particle from the cavity, thus changing
the absorption of the composite system from $0.095$ to $0.150$ (see
left vertical dashed line in Fig.~\ref{fig4}). The opposite happens
for the second resonance mode located at higher wavelengths: the
force is directed downwards when the rod is placed close to the top
of the cavity (solid curves), thus pushing the particle downwards.
For an operating wavelength $\lambda_2=1030~\mathrm{nm}$ (see right
vertical dashed line in Fig.~\ref{fig4}) the trapping of the
particle becomes possible inside the cavity.

\begin{figure}[t]
\centerline{\includegraphics*[width=8.5cm]{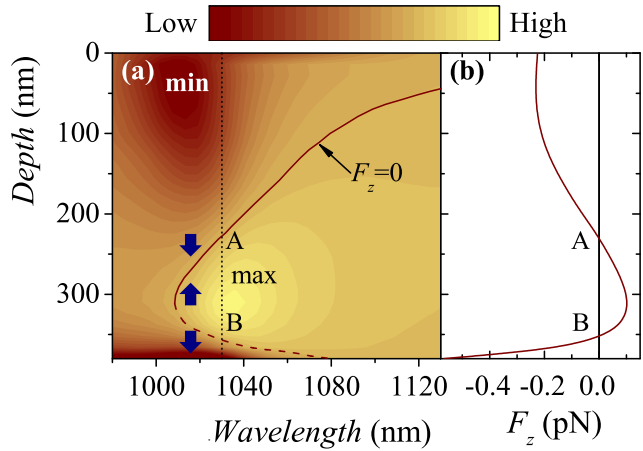}}
\caption{(color online) {\bf (a)} Vertical force acting on the rod
as a function of light wavelength and particle position (depth)
below the cavity opening. The external light intensity is
$1~\mathrm{mW/\mu m^2}$. Solid and dashed curves denote stable and
unstable equilibrium (zero-force) points, respectively. The blue
arrows visualize the direction of the optical force in the vicinity
of the wavelength $1030~\mathrm{nm}$, indicated by the dotted
vertical line, for which the detailed form of the force is presented
in {\bf (b)}. Points $\mathrm{A}$ and $\mathrm{B}$ show stable and
unstable trapping positions inside the cavity, respectively.}
\label{fig5}
\end{figure}

This is demonstrated in more detail in Fig.~\ref{fig5}. The $z$
component of the force acting on the rod placed at different depths
inside the cavity is shown for different wavelengths in
Fig.~\ref{fig5}(a) (light intensity equal to $1~\mathrm{mW/\mu
m^2}$). The force surface-map exhibits stable and unstable
equilibrium (zero-force) points, denoted by solid and dashed curves
in the figure, respectively. Stable trapping of the rod at a given
depth inside the cavity occurs when the force changes direction from
positive to negative when moving from below to above that depth.
Consequently, this will be the case for wavelengths in the
$1020-1040~\mathrm{nm}$ range and for points lying on the
stable-equilibrium line [solid curve in Fig.~\ref{fig5}(a)]. As an
example, we choose the wavelength of $1030~\mathrm{nm}$. The
corresponding depth-profile force is depicted in Fig.~\ref{fig5}(b).
For this specific wavelength, point A (depth$\sim 220~\mathrm{nm}$)
represents a trapping position for the rod, while point B
(depth$\sim 350~\mathrm{nm}$) is an anti-trapping position.

In conclusion, we have shown that composite systems formed by gold
nanoparticles trapped inside water-filled gold nanocavities exhibit
large variations in optical absorption as a function of the position
of the particle inside the cavity. Optical forces acting on the
particle have been shown to be sufficiently large to move the
particles around, and in particular, positions and illumination
wavelengths have been identified that lead to stable particle
trapping. The combined cavity-nanoparticle system is thus tunable
via the wavelength of a {\it pump} light source, which triggers
dramatic changes in the reflectance spectrum experienced by a second
{\it probing} light source. Our findings are thus opening a new path
towards all-optical switching.

\begin{acknowledgments}
This work was supported by the Spanish MEC (NAN2004-08843-C05-05 and
MAT2007-66050) and by the EU-FP6 (NMP4-2006-016881
"SPANS").
\end{acknowledgments}


\begin{thebibliography}{22}
\bibitem{forces} At room temperature $\mathrm{T}\sim300~\mathrm{K}$ and for
  distances $D$ of the order of $100~\mathrm{nm}$ (i.e., of the order of the
  cavity size), the thermal force can be roughly estimated from $F_{th}\sim K_B
  \mathrm{T}/D$ to be of the order of $40~\mathrm{fN}$, where $K_B$ is the Boltzman
  constant.
\end{thebibliography}
\end{document}